\documentclass{appolb}
\usepackage{epsfig}
\usepackage{amssymb}
\usepackage{amsmath}

\begin{document}
\title{Physics Prospects at FAIR
\thanks{Presented at the international conference on Strangeness in Quark Matter 2011}%
}
\author{M.~Bleicher$^{1}$, M.~Nahrgang$^{1,2}$, J.~Steinheimer$^{1}$, Pedro Bicudo$^{3}$
\address{$^{1}$ Frankfurt Institute for Advanced Studies (FIAS), Ruth-Moufang-Str.~1, and Institut f\"ur Theoretische Physik, Johann Wolfgang Goethe University, 60438 Frankfurt am Main, Germany}
\address{$^{2}$ SUBATECH, UMR 6457, Universit\'e de Nantes, Ecole des Mines de Nantes,
IN2P3/CRNS. 4 rue Alfred Kastler, 44307 Nantes cedex 3, France}
\address{$^3$ CFTP, Departamento de Fisica, Instituto Superior Tecnico (Universidade
Tecnica de Lisboa), Av. Rovisco Pais, 1049-001 Lisboa, Portugal}
}
\maketitle
\begin{abstract}
We review the physics potential at FAIR in the light of the existing data of the RHIC-BES program and the NA49/NA61 beam and system size scan. Special emphasize will be put on the potential of fluctuations, as well as dilepton observables.
\end{abstract}
  
\section{Introduction}
The Facility for Antiproton and Ion Research, FAIR \cite{bib1a,bib1b,bib1c}, will provide an extensive range of particle beams from protons and antiprotons to ion beams of all chemical elements up to the heaviest one, uranium, with in many respects world record intensities. As a joint effort of several countries the new facility builds, and substantially expands, on the present accelerator system at GSI, both in its research goals and its technical possibilities. Compared to the present GSI facility, an increase of a factor of 100 in primary beam intensities, and up to a factor of 10000 in secondary radioactive beam intensities, will be a technical property of the new facility.
The main thrust of FAIR research focuses on the structure and evolution of matter on both a microscopic and on a cosmic scale. The approved FAIR research programme embraces 14 experiments, which form the four scientific pillars of FAIR and offers a large variety of unprecedented forefront research in hadron, nuclear, atomic and plasma physics as well as applied sciences. Already today, over 2500 scientists and engineers are involved in the design and preparation of the FAIR experiments. They are organized in the experimental collaborations APPA, CBM, NuSTAR, and PANDA.
The CBM/HADES experiment is of particular interest for the understanding of highly compressed nuclear matter and its relevance for understanding fundamental aspects of the strong interaction. HADES \cite{bib3a,bib3b,bib3c} and CBM \cite{bib3d,bib3e} at SIS100/300 will explore the QCD phase diagram in the region of very high baryon densities and moderate temperatures. This approach includes the study of the nuclear matter equation-of-state, the search for new forms of matter, the search for the predicted first order phase transition between hadronic and partonic matter, the QCD critical endpoint, and the chiral phase transition, which is related to the origin of hadron masses. It is intended to perform comprehensive measurements of hadrons, electrons, muons and photons created in collisions of heavy nuclei proton--nucleus, and proton--proton collisions at different beam energies. Most of the rare probes like lepton pairs, multi-strange hyperons and charm will be measured for the first time in the FAIR energy range. 

\section{Dileptons}
Dileptons represent a penetrating probe of the 
hot and dense nuclear matter created in heavy ion collisions at the CBM experiment at the FAIR facility.  
The analysis of the electromagnetic response of the dense and hot medium 
is tightly connected to the investigation of the in-medium modification 
of the vector meson properties. Vector mesons are ideally suited for this 
exploration, because they can directly decay into a 
lepton-antilepton pair. One therefore aims to infer information on the 
modifications induced by the medium on specific properties of the vector 
meson, such as its mass and/or its width, from the invariant mass dilepton 
spectra.
In this work, we present a consistent calculation 
of the dilepton production at SPS energy within a model which attempts to take 
into account both the complexity of the dilepton rate in hot a dense medium 
as well as the complexity of the pre-, post-, and equilibrium 
heavy-ion dynamics. 
The latter is modelled with an integrated 
Boltzmann+hydrodynamics hybrid approach based on the Ultrarelativistic 
Quantum Molecular Dynamics (UrQMD) transport model with an intermediate 
(3+1) dimensional ideal hydrodynamic stage \cite{Petersen:2008dd}. During the locally equilibrated hydrodynamical stage, dimuon  emission is calculated locally in space-time according to the expression for the thermal equilibrium rate of dilepton emission 
per four-volume and four-momentum from a bath at temperature $T$ and baryon chemical potential $\mu_B$.
During the local equilibrium phase, the 
radiation rate of the strongly interacting medium is standardly 
modelled using the vector meson dominance model and related to the spectral 
properties of the light vector mesons, with the $\rho$ meson having the 
dominant role \cite{Gale:1990pn,Rapp:1999ej,Ruppert:2007cr,vanHees:2007th}.
In-medium modifications of the $\rho$-meson spectral function due to 
scattering from hadrons in the heat bath are properly included in the model.
Two additional sources of thermal radiation, namely  emission 
from four-pion annihilation processes and from a thermalized partonic 
phase are included as well.
As an input for the hydrodynamical part of the evolution we employ 
an equation of state in line with lattice data that 
follows from coupling the Polyakov loop to a chiral 
hadronic flavor-SU(3) model \cite{Steinheimer:2010ib}.

\begin{figure*}[t]
\includegraphics[width=.45\textwidth,angle=-90]{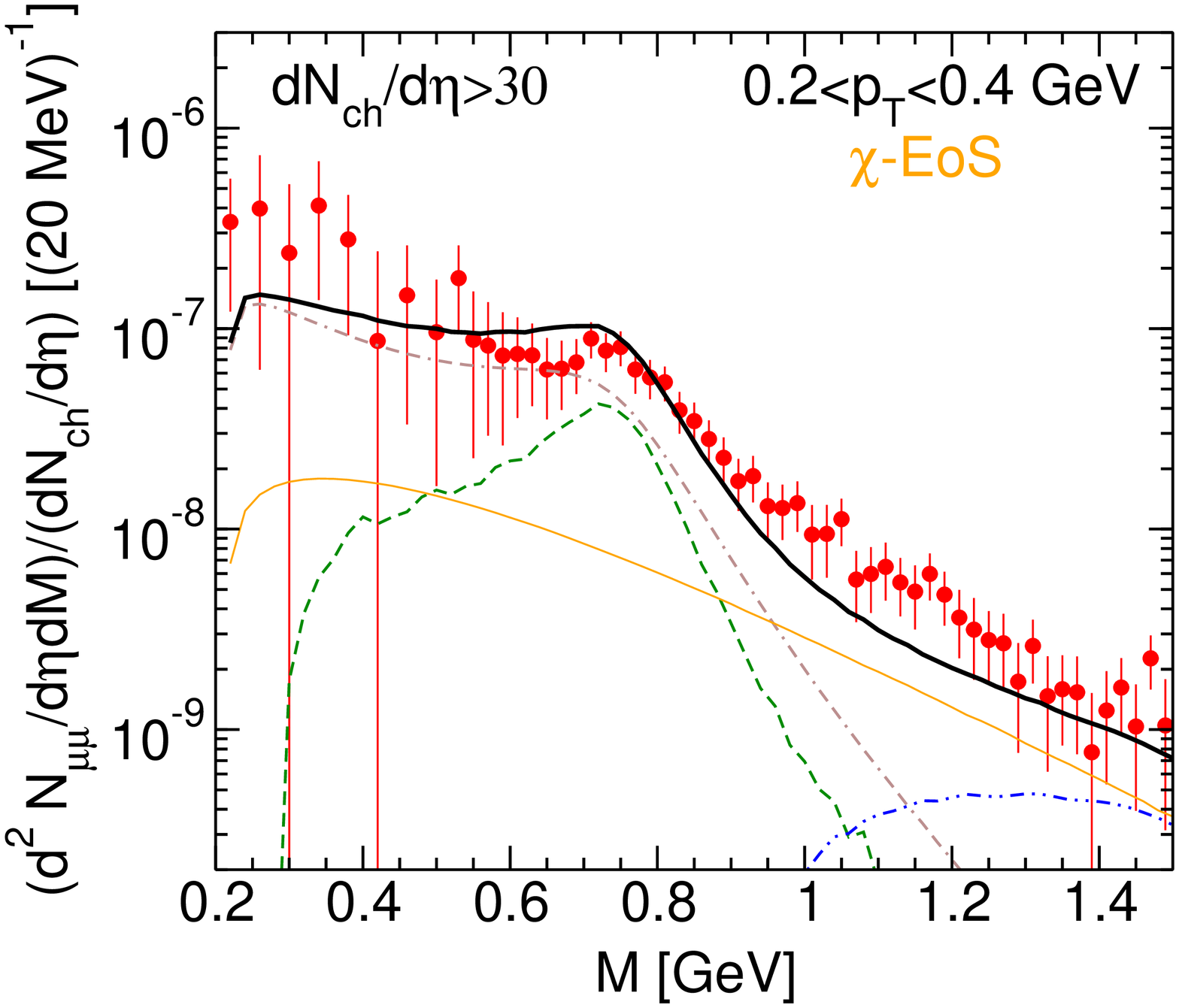}
\includegraphics[width=.45\textwidth,angle=-90]{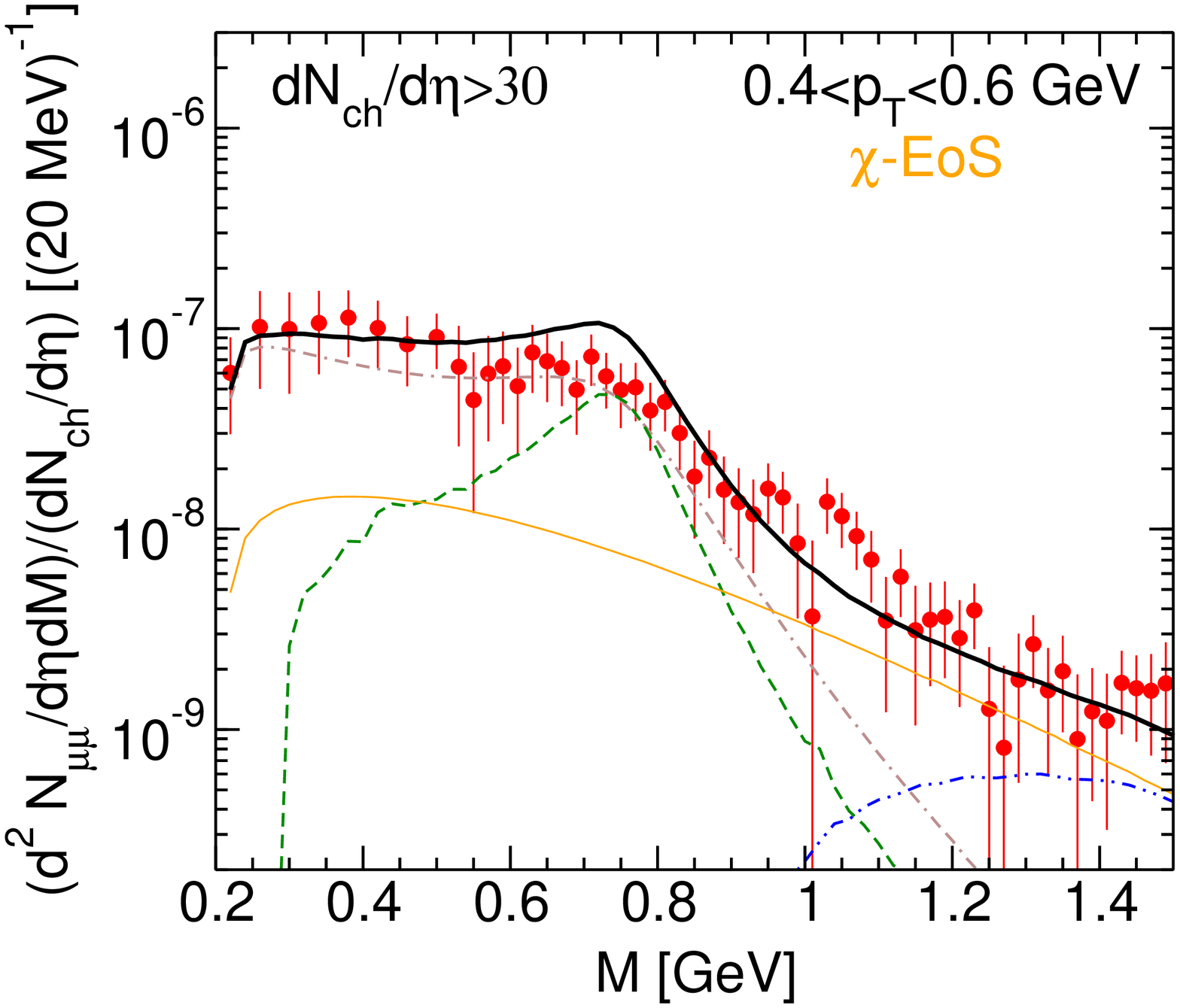}
\caption{Acceptance-corrected invariant mass spectra of the 
excess dimuons in In-In collisions at 158$A$ GeV for various  bins of 
transverse pair momenta. 
The individual contributions arise from in-medium modified $\rho$ mesons 
(dotted-dashed line), $4\pi$ annihilation (double dotted-dashed line), 
quark-antiquark annihilation in the QGP (thin full line) and 
cascade $\rho$ mesons (dashed line). 
The sum of the various contributions is depicted by the thick full line.
Experimental data from Ref. \cite{Arnaldi:2008fw}. \label{fig1}}
\end{figure*}

The results we show for the dilepton spectra in In+In collisions at $E_{lab}=160 A$ GeV will be compared to fully acceptance corrected  NA60 data \cite{Arnaldi:2008fw}. The data correspond to nearly minimum bias collisions, selecting events with a charged particle density $dN_{ch}/d\eta$$>$30.
In Fig. \ref{fig1} we show results for the invariant mass spectra of 
the excess dimuons in two slices in the transverse momentum of the 
dilepton pair $p_T$ when adopting a sudden freezeout approximation. The theoretical spectra are normalized to the corresponding average number of charged particles in an interval of one unit of rapidity around mid-rapidity. Results on more bins in transverse momentum as well as a detailed discussion on the applied model and results can be found in \cite{arXiv:1102.4574}. Of particular interest is that in the intermediate mass region, 1$<$$M$$<$1.5 GeV, we find that emission from the QGP accounts for 
about half of the total radiation. The remaining half is filled by the 
considered hadronic sources. 
The $4\pi$ annihilation alone is comparable to the QGP emission only for 
$M$$>$1.4 GeV. This offers the possibility to even quantitatively pin down the QGP contribution to the dilepton spectra and consequentely to the active degrees of freedom for the SIS beam energies.

\section{Fluctuations induced by a phase transition}
At larger baryochemical potential, as achieved at FAIR, a first order phase transition is expected from model studies \cite{Bleicher:1998wu,Scavenius:2000qd,Ratti:2005jh,arXiv:0704.3234}. Interesting observables could here be based on the growth of fluctuations due to the nonequilibrium effect of supercooling leading to nucleation and spinodal decomposition \cite{Csernai:1992tj,Mishustin:1998eq,Randrup:2010ax,Chomaz:2003dz}.
At zero baryochemical potential the nature of the phase transition of QCD is well understood from lattice QCD calculations, which show that it is an analytic crossover \cite{Aoki:2006we}. As a consequence there must be a critical point, which terminates the line of first order phase transitions. In equilibrium systems fluctuations and correlations of the order parameter diverge at the critical point. Coupling particles to the sigma field, the order parameter of chiral symmetry, leads to a nonmonotonic behaviour in fluctuations of net-charge or net-baryon number multiplicities  \cite{Stephanov:1998dy,Stephanov:1999zu}. The key ingredient is the correlation length which becomes infinite in a system at a critical point. In a realistic evolution of a heavy-ion collision, however, the growth of the correlation length is limited by the size of the system and by the finite time, which the dynamic systems spends at a critical point. Relaxation times also become infinite at the critical point, a phenomenon called critical slowing down. Even if the system is in equilibrium above the critical point it is necessarily driven out of equilibrium by passing trough the critical point. Assuming a phenomenological time evolution of the correlation length with parameters from the $3$d Ising universality class it was found that the correlation length does not grow beyond $2-3$ fm \cite{Berdnikov:1999ph}.
The explicit propagation of fluctuations coupled to a dynamic model is a necessary step towards understanding the QCD phase diagram from heavy-ion collision experiments. Here we present results from a recently developped extention of chiral fluid dynamics  \cite{Mishustin:1998yc,Paech:2003fe}, which self-consistently includes the nonequilibrium propagation of the fluctuation of the order parameter of chiral symmetry, the sigma field \cite{Nahrgang:2011mg,Nahrgang:2011ll,arXiv:1105.1962}. It is coupled to a fluid dynamic expansion, where the fluid is made out of quarks and antiquarks and acts as a locally equilibrated heat bath.
Due to the interaction with the quark fluid the sigma fields is damped. This is taken into account by dissipative terms in the Langevin equation of motion of the chiral fields
\begin{equation}
 \partial_\mu\partial^\mu\sigma+\frac{\delta U}{\delta\sigma}+\frac{\delta \Omega_{\bar qq}}{\delta\sigma}+\eta\partial_t \sigma=\xi\, .
\label{eq:equi_langevineq}
\end{equation}
It contains a classical Mexican hat potential $U$, the quark contribution to the thermodynamic potential $\Omega_{\bar qq}$ to one-loop level, the damping coefficient $\eta$ and the stochastic noise field $\xi$. The dynamics of the quarks is reduced to the propagation of densities according to energy-momentum conservation, i.e. the equations of relativistic fluid dynamics
\begin{equation}
\partial_\mu T^{\mu\nu}=S^\nu\, ,
\label{eq:fluidT}
\end{equation}
where the source term $S^\nu$ accounts for the energy-momentum exchange between the fluid and the field.

\begin{figure}
  \centering
  \includegraphics[width=.45\textwidth]{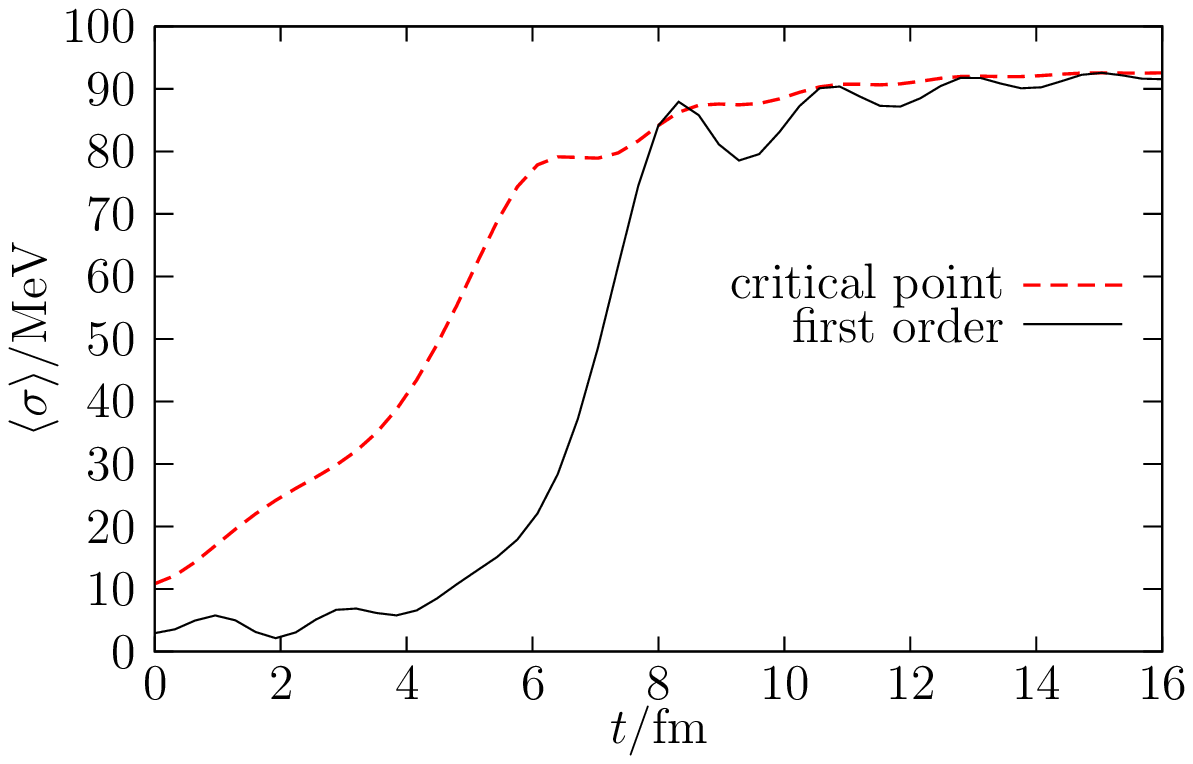}
  \includegraphics[width=.45\textwidth]{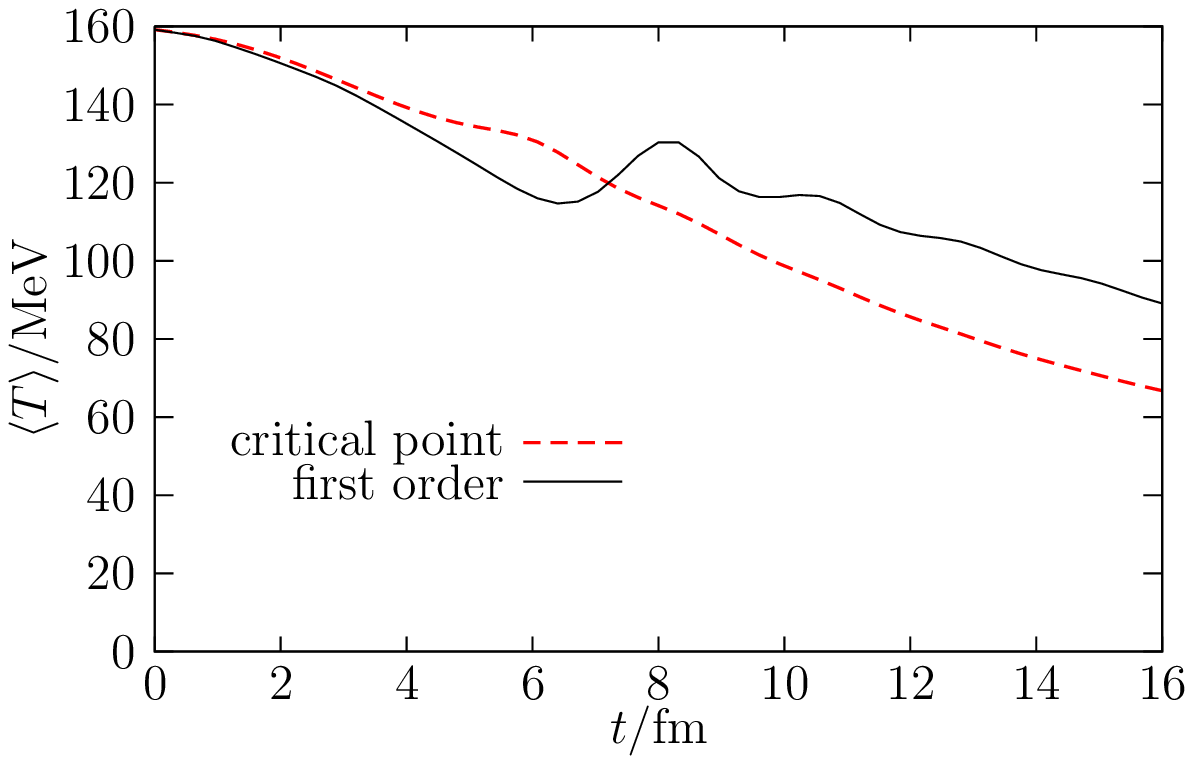}
\caption{The average of the sigma field (left) and of the temperature (right) for $\eta=2.2$/fm and both phase transition scenarios.}
  \label{fig:cf2_hotregionsigmaeta22}
 \end{figure}

Here we apply a constant value of $\eta=2.2$/fm \cite{Biro:1997va}. The time evolution of the average value of the sigma field $\langle\sigma\rangle$ and the average temperature $\langle T\rangle$ is shown in figure \ref{fig:cf2_hotregionsigmaeta22}. The average is taken over an initially hot and dense sphere with radius $r=3$~fm. The phase transition temperature in a critical point scenario $T_c=139.88$~MeV is crossed at around $t=5$~fm, after which the slightly enhanced fluctuations fall off. The average sigma field smoothly relaxes towards its vacuum value. As the phase transition temperature of the first order phase transition, $T_c=123.27$~MeV, is lower than at a critical point the system relaxes later but the relaxational process itself is faster in a first order phase transition scenario. The vacuum value is reached around the same time, but the average sigma field shows strong oscillations and the fluctuations are enhanced in the first order phase transition scenario. 

When the first order phase transition temperature is reached after $t\simeq 5$~fm large parts of the system are still in the chirally broken phase as the average value of the sigma field is still $\langle\sigma\rangle\lesssim10$~MeV. These large deviations of the sigma field from its equilibrium value is the nonequilibrium effect of supercooling.

Due to the steep curvature in the effective potential experienced by the system once the barrier is overcome, the potential energy is transformed effectively in kinetic energy, which leads to the dissipation of energy via $\eta(\partial_t\sigma)^2$ in the source term. In figure \ref{fig:cf2_hotregionsigmaeta22} we can clearly observe the reheating effect at the first order phase transition. Between $t=7$~fm and $t=9$~fm the system is reheated from $T\simeq 118$~MeV below $T_c$ to $T\simeq 125$~MeV above $T_c$, followed by a subsequent cooling. Thus, the reheating causes the system to cross the phase transition two more times, once in the reverse direction from the low temperature phase to the high temperature phase around $t=8$~fm and again at around  $t=9$~fm. This contributes to a slower relaxation of the average sigma field.

The effective potential with a critical point is very flat at the transition temperature and reheating is not observed. Instead the cooling is slightly decelerated as seen in figure \ref{fig:cf2_hotregionsigmaeta22} between $t=5$~fm and $t=6$~fm.
 
\begin{figure}
  \centering
  \includegraphics[width=.75\textwidth]{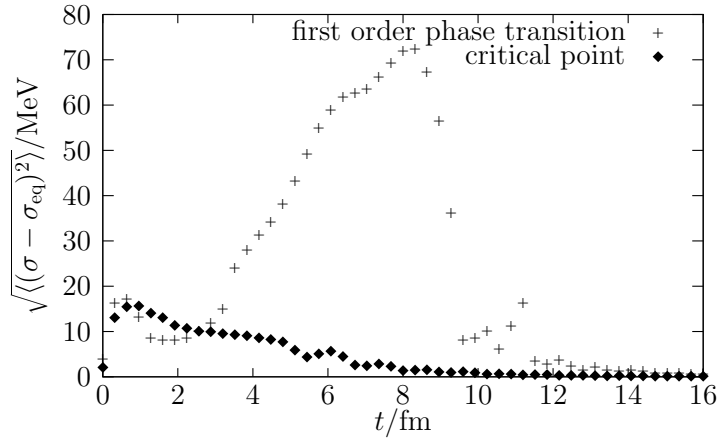}
\caption{Time evolution of the deviation of the sigma field from the thermal equilibrium value for a scenario with a first order phase transition and with a critical point.}
  \label{fig:flucgleich}
 \end{figure}

The sigma field is locally initialized in equilibrium with the quark fluid. The subsequent time evolution of the averaged sigma fluctuations around the thermal equilibrium value plotted in figure \ref{fig:flucgleich}. In a critical point scenario the fluctuations are slightly increased in the beginning of the expansion due to the dynamics of the system. As expected we do not observe an increase at the critical temperature. For the first order phase transition this is very different. Due to the effect of supercooling we find a large enhancement of nonequilibrium fluctuations after the phase transition temperature is reached. During relaxation parts of the system are reheated to temperatures above the phase transition which leads to a second increase in the fluctuations. 
Extentions of this model to include the deconfinement phase transition are presented in \cite{herold}.

\section{Summary}
We presented results where for the first time dilepton emission is 
studied within a macro+micro hybrid approach.
We found that three regions can be identified in the dilepton invariant spectra. 
The very low mass region of the spectrum is dominated by thermal 
radiation, the region around the $\rho$ meson peak is 
dominated by late stage cascade dilepton emission and the 
intermediate region receives both contributions from hadronic and 
QGP emission, with the QGP accounting for about half of the total emission. Such studies will be important to be able to connect dilepton measurements at the CBM experiment with the phase transition to the QGP.

In the following we have shown results on the explicit propagation of fluctuations at the chiral phase transition within a dynamic model of heavy-ion collisions. During the expansion the system cools and crosses the phase transition where nonequilibrium effects occur. These are expecially dominant in a first order phase transition leading to supercooling and reheating of the entire system. Since nonequilibrium fluctuations are large at the first order phase transition and persist for some time below the phase transition there is a potential of observing the first order phase transition in heavy-ion collisions at the CBM experiment.

This work was supported by the Hessian LOEWE initiative Helmholtz International Center for FAIR and used computational resources provided by the (L)CSC at Frankfurt.

\end{document}